\documentclass[prb,aps,twocolumn,showpacs]{revtex4}
\usepackage{amsmath,amssymb,amsfonts,float}
\usepackage[dvips]{graphicx,color}    
\usepackage{times}       
 
\newcommand{\bolM}{\text{\bf M}}

\newcommand{\bolk}{\mathbf{k}}

\newcommand{\bolx}{\mathbf{x}}

\newcommand{\VEV}[1]{\langle #1 \rangle}  

\newsavebox{\dotdot}
\savebox{\dotdot}[3mm]{\shortstack{\circle*{0.8}\\ \\ \circle*{0.8}}}

\begin{document}
\title{Topological charge pumping effect by the magnetization dynamics on the Surface of Three-Dimensional Topological Insulators}
\author{Hiroaki~T.~Ueda$^1$, Akihito~Takeuchi$^1$, Gen~Tatara$^1$, Takehito~Yokoyama$^2$\\
$^1$Department of Physics, Tokyo Metropolitan University, Hachioji, Tokyo 192-0397, Japan\\
$^2$Department of Physics, Tokyo Institute of Technology, Tokyo 152-8551, Japan}
\begin{abstract}
We discuss a current dynamics on the surface of a 3-dimensional topological insulator induced by magnetization precession of an attached ferromagnet. 
It is found that the magnetization dynamics generates a direct charge current when the precession axis is within the surface plane. 
This rectification effect is due to a quantum anomaly and is topologically protected. 
The robustness of the rectification effect against first-varying exchange field and impurities is confirmed by explicit calculation. 
\end{abstract}            
\maketitle
\section{introduction}
A topologically-classified phase has universal properties which are robust against perturbation. 
Topological aspect emerges regardless of classical and quantum systems, 
and it is expected to be a useful tool to develop devices using nontrivial quantum phenomena. 
Recently, a topological insulator (TI), which is a gapped insulator in bulk and has a topologically characterized gapless edge state, has been drawing attention\cite{Moore,Fu,ME,Hasan,review} and the various exotic phenomena have been reported, such as the surface quantum Hall effect and the realization of the monopole dynamics.\cite{review,monopole} 
For a TI in three space dimensions (3D), Qi {\it et. al.} 
argued that the surface state is described by the massive (gapped) (2+1)-dimensional Dirac fermion and non-trivial topological magnetoelectric effect appears when the exchange field is present.\cite{ME} 
The low energy response to the electromagnetic field 
was shown to be understood by the topological mass (TM) term (Chern-Simons term)\cite{P-anomaly1,P-anomaly2,Tmass,TKNN,Coleman-Hill}: 
\begin{equation}
S_{\text{TM}}(A)=\frac{e^2}{8\pi}\frac{m}{|m|}\int d^3x \epsilon^{\mu\nu\rho}A_\mu \partial_\nu A_\rho\ .
\end{equation}
where $d^3x=dtd^2x$, $m$ is the mass of the Dirac fermion and $A$ is the $U(1)$ gauge field. 

One of the interesting aspects of TM is the quantum anomaly, where quantum fluctuations break the symmetry in the original Lagrangian.
In fact, TM breaks the parity symmetry which exists when Dirac mass is zero: the  so-called parity anomaly. 
Originally, the quantum anomaly was discussed in the context of the chiral anomaly in the decay of the $\pi$ meson, which has been experimentally well confirmed.\cite{anomaly,PeskinSchroeder} 
Generally, the anomaly emerges from the regularization of the 
ultraviolet divergent diagram in the diagrammatic expansion. 
In condensed matter physics, it is known that the graphene exhibits exotic physics because of TM, such as the quantum Hall effect\cite{Haldane} and the quantum spin Hall effect.\cite{KaneMele} It is important that the system is described by the Dirac Lagrangian, where the Green's function behaves as $O(1/|k|)$ for $k\rightarrow\infty$ ($k$ is the momentum of the Dirac particle), and thus the ultraviolet divergence is likely to occur in contrast to the nonrelativistic case with the Green's function $\propto O(1/k^2)$. In the study on the TI by the diagrammatic expansion, therefore, the divergent diagram should be treated carefully. 
Since the ultraviolet divergence is inevitable in calculating physical quantities related to the electromagnetic response of the TI,
it is a fundamental issue how to regularize the divergence. 

Another interesting aspect of TM is the quantization of physical observables. 
In diagrammatic studies, TM appears in the bare one-loop Feynman diagram in the self energy of the electromagnetic $U(1)$ gauge boson.\cite{P-anomaly1} 
As shown by the Coleman and Hill, the TM obtained in the 1-loop diagram is exact, and the correction from scalar, spinor, vector and gauge fields is irrelevant.\cite{Coleman-Hill} 
The coefficient of the TM is quantized, and the mechanism can be understood from the viewpoint of the Berry's phase. 
In the band insulator, the correspondence between the TM and the Berry's phase is explicitly seen from the Hall conductance obtained by the Kubo formula.\cite{TKNN,Zhang} 
The Berry phase represents the topological number and ensures the topological robustness of the TM.\cite{review,XGWenText} 

The robustness of the TM against impurities, as it is topologically expected, is crucially important in obtaining experimental evidence of the TM.
This robustness can be understood by the Coleman-Hill theorem in the gapped Dirac systems. 
In diagrammatic calculations, after averaging the positions of impurities, the impurities can be viewed as the chargeless scalar field and thus the resulting diagrams satisfy the Ward-Takahashi identity and the momentum conservation. 
Hence, as long as the pure system free from the impurities is insulating and the fermion gap opens, the Coleman-Hill theorem for scalar fields can be applied and the TM persists against impurities. 
Recently, Nomura and Nagaosa studied the effect of the various type of the impurities  which can break the time reversal symmetry and can alter the mass gap on the surface state of the TI.\cite{Nomura-Nagaosa2} 
They confirmed that the impurities does not affect the TM within the linear response theory. 
Moreover, they suggested that the metallic state is localized by the magnetic impurities and, even in that case, the magnetoelectric response can be described by the TM. 

The gapped surface state of the TI where the TM characterizes the electromagnetic response is realized by depositing the ferromagnet (FM) on the surface of the TI.\cite{review,ME}
In this system, the magnetization of the FM plays the same role as the $U(1)$ gauge field and the Dirac mass. 
By using this property, various spintronics phenomena have been predicted.\cite{Inverse,Tse,Maciejko,Yokoyama1,Yokoyama2,Burkov,Nomura-Nagaosa,Yokoyama3,Yokoyama4,Mahfouzi,Tserkovnyak}
For example, it has been proposed that the current applied on the surface of the TI can flip the magnetization.\cite{Inverse,Yokoyama2,Yokoyama4} 
Nomura and Nagaosa showed that the magnetic textures of FM show a nontrivial dynamics when the electric field is applied, such as the motion of the magnetic domain wall.\cite{Nomura-Nagaosa} 
These new effects are expected to lead to new spintronics devices. 

\begin{figure}[htbp]
\begin{center}
\includegraphics[scale=0.6]{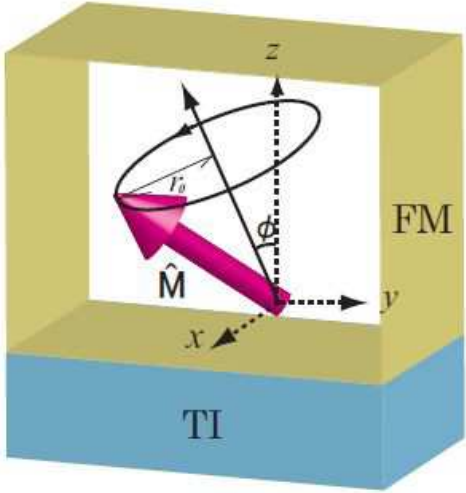}
\caption{(color online) 
A ferromagnet (FM) is attached on the surface of the TI. The direction of the magnetization $\hat{\bolM}$ of the FM is externally controlled. The current is pumped on the surface of TI by the rotation of the magnetization. The precession axis is assumed in the $y$-$z$ plane.
\label{Fig;rotM}} 
\end{center}
\end{figure}

In this paper, we demonstrate that the surface of the TI exhibits the rectification effect because of the parity anomaly when the current is pumped by the magnetization dynamics of FM (see Fig.~\ref{Fig;rotM}). 
The rectification effect occurs in the adiabatic limit, i.e., when the dynamics of the magnetization is slow, and the validity of the adiabatic approximation is studied diagrammatically. 
To avoid the divergence, we adopt the dimensional regularization developed by 't Hooft and Veltman.\cite{D-regularization} 
The dimensional regularization is one of the simplest regulators which ensures the gauge invariance of the theory, and is generally applicable. 
In fact, this regularization is successfully applied to the study on the conductivity of the graphene.\cite{grapheneDR} 

We also study the effect of the impurities by using the perturbation theory when the energy gap of the surface state opens. 
Although the robustness of the TM is ensured by the Coleman-Hill theorem, 
it may be important to see the robustness by an explicit diagrammatic calculation. It turns out that the first-order result gives the correct quantized TM, regardless of the impurity concentration.


\section{rectification effect}
\label{Sec:rect}
In this section, we study the current driven by the magnetization precession in a junction of TI and FM. 
The surface state of the TI is described by the (2+1)D massless Dirac Hamiltonian. 
When a FM is attached, the spin-exchange interaction appears and the Hamiltonian is given by\cite{review}
\begin{equation}
H=\int d^2x\ v_{\text{F}}\psi^\dagger(\bolx)(p_y\sigma_x-p_x\sigma_y)\psi(\bolx)+\Delta \psi^\dagger(\bolx)(\hat{\bolM}\cdot\mbox{\boldmath $\sigma$}) \psi(\bolx)\ ,
\end{equation}
where $v_{\text{F}}$ is the Fermi velocity, $\bolx=(x,y),\ \mbox{\boldmath $\sigma$}=(\sigma^x,\sigma^y,\sigma^z)$, the direction of the magnetization is represented by a unit vector $\hat{\bolM}=(\hat{M}_x,\hat{M}_y,\hat{M}_z)$ and $\Delta$ is the spin exchange coupling energy between the surface fermion and the local magnetization of the ferromagnet.
We ignore the effect of the electoromagnetic field 
\footnote{The coupling between the surface electron and the external electromagnetic field $\tilde{A}_\mu$ is given by the $\partial_{\mu}\rightarrow \partial_{\mu}-ie\tilde{A}_{\mu}/\hbar c$ and the Zeeman coupling $g\mu_B {\bf B}\cdot \mbox{\boldmath $\sigma$}$. 
The electromagnetic response is given by 
${\bf \rho}=\frac{e^2}{2h}\frac{m}{|m|}B_z$ and $j_i=\frac{e^2}{2h}\frac{m}{|m|}\epsilon_{ij}E_j$ for $i,j=(x,y)$.}
resulting from the magnetization precession since the magnetic field itself does not affect the charge current, and the electric field, which may be proportional to a system size,
\footnote{
Naively, the induced electric field at the edge of the FM surface is given by $E_{i}\sim r\omega_0 M_z/2$, where $i=(x,y)$ and $r,\ \omega_0,\ M_z$ are respectively the system size, frequency of the precession and $z$ component of the magnetization of the FM. }
can be neglected in a small-size system. 
We assume that the Fermi level is at the Dirac point. 
Then, the Lagrangian is simplified as
\begin{equation}
L=\int v_{\text{F}}d^2x\ \bar{\psi}(\bolx)(i\hbar \gamma^\mu D_\mu -mv_{\text{F}})\psi(\bolx) \
\end{equation}
where $\mu=0,1,2$ is the space-time index and
\begin{equation}
\begin{split}
D_\mu&=\partial_\mu-i\frac{eA_\mu}{\hbar v_{\text{F}}},\ A_\mu=(0,\frac{\Delta \hat{M}_y}{e},-\frac{\Delta \hat{M}_x}{e})\ ,\\
\gamma^\mu&=(\frac{\sigma^z}{v_{\text{F}}},i\sigma^x,i\sigma^y),\ \bar{\psi}=\psi^\dagger\sigma^z,\ m=\frac{\Delta \hat{M}_z}{v_{\text{F}}^{2}}\\
x^\mu&=(t,\bolx),\ g_{\mu\nu}=\text{diag}(v_{\text{F}}^2,-1,-1)\ .
\end{split}
\end{equation}
Here, $\gamma^\mu$ are the $2\times 2$ matrices and satisfy the Dirac algebra $\{\gamma^\mu,\gamma^\nu\}=2g^{\mu\nu}\times {\bf 1}_{2\times 2}$ (${\bf 1}_{2\times 2}$ is the $2\times 2$ unit matrix). 
The Lagrangian is thus equivalent to the conventional Dirac fermion of the charge $e(<0)$ coupled with the electromagnetic field.  In the early 1980's, it was found that the current is described by the TM\cite{P-anomaly1,P-anomaly2}:
\begin{equation}
\VEV{j^\mu}= \frac{e^2m}{4\pi\hbar v_{\text{F}}|m|}\epsilon^{\mu\nu\rho}\partial_\nu A_\rho+O(\frac{e^2\partial_\nu\partial^\nu A}{\sqrt{|g_{\mu\mu}|}mv_{\text{F}}})
\label{j}
\end{equation}
where $j^\mu=v_{\text{F}}e\bar{\psi}(x)\gamma^\mu\psi(x)$, the summation over $\nu$, $\rho$ is implied and 
$|\hbar \partial_i A /A|,|\hbar v_{\text{F}}^{-1} \partial_t A /A| \ll mv_{\text{F}}$. 
In terms of the magnetization, the current is given as discussed in Ref.~\onlinecite{Nomura-Nagaosa} by
\begin{equation}
\begin{split}
\VEV{j^0(x)}&=\rho(x)= \frac{e\Delta m}{4\pi\hbar v_{\text{F}}|m|}(\partial_x \hat{M}_x+\partial_y \hat{M}_y)\ ,\\
\VEV{j^i(x)}&= -\frac{e\Delta m}{4\pi\hbar v_{\text{F}}|m|}\partial_t \hat{M}_i\ ,
\end{split}
\label{current}
\end{equation}
where $i=(x,y)$. 

These results are for the case of constant $m$, but they remain correct as long as the adiabatic condition is satisfied, namely, if the mass of the Dirac particle, $m$, is larger than the frequency of the magnetization
multiplied by $\hbar\omega_0/v_{\text{F}}^2$. 
The validity of eq.~(\ref{current}) in the adiabatic regime will be demonstrated in Sec.~\ref{adiabatic}, and we will proceed here based on eq.~(\ref{current}). 
From eq.~(\ref{current}), we see that $\VEV{j^0}$ and $\VEV{j^i}$ exhibit anomalous behavior when $m$ crosses zero because of the coefficient $m/|m|=\text{sign}(m)$. 
In fact, if $\partial_t \hat{M_i}$ is finite when $m=0$, 
$\VEV{j^i}$ exhibits a sudden jump. 
If $\partial_t \hat{M}_i$ changes sign at $m=0$, 
the sign of the current is the same for both $m>0$ and $m<0$, and the generated current is rectified due to the anomaly.
 
Let us look into these behaviors in detail in the case of precessing magnetization 
with the angular frequency $\omega_0$. 
We choose the angle $\phi$ as the angle of the precession axis measured 
from the $z$-axis (Fig.~\ref{Fig;rotM}), and the radius of the precession 
is denoted by $r_0 (\leq 1)$.
The magnetization is then represented as
\begin{equation}
\begin{split}
\hat{\bolM}=
\left(
\begin{array}{c}
r_0\sin \omega_0 t\\
r_0\cos\phi\cos \omega_0 t+\sqrt{1-r_0^2}\sin\phi\\
-r_0\sin\phi \cos\omega_0 t+\sqrt{1-r_0^2}\cos\phi
\end{array}
\right)\ .
\end{split}
\label{eqM}
\end{equation}
The induced current is given by $\VEV{j^0}=0$ and
\begin{equation}
\begin{split}
\VEV{j^x}&=\frac{e\Delta}{4\pi\hbar v_{\text{F}}}\frac{\hat{M_z}}{|\hat{M_z}|}r_0\omega_0\cos \omega_0 t,\\
\VEV{j^y}&=-\frac{e\Delta}{4\pi\hbar v_{\text{F}}}\frac{\hat{M_z}}{|\hat{M_z}|}r_0\omega_0\cos\phi\sin \omega_0 t\ .
\end{split}
\end{equation}

\begin{figure}[htbp]
\begin{center}
\includegraphics[scale=0.5]{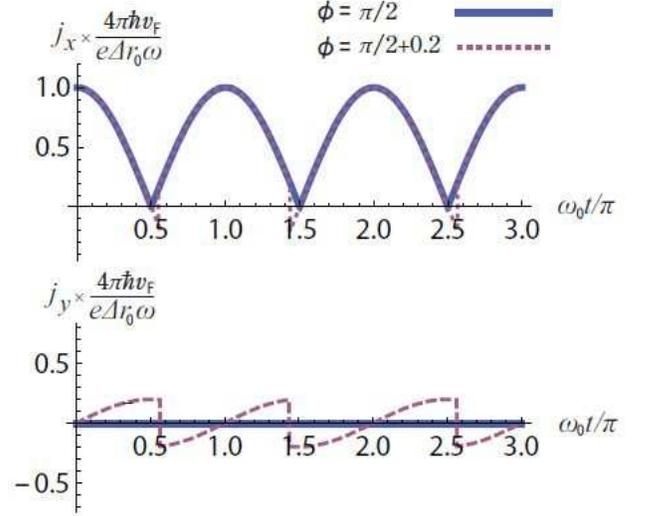}
\caption{(color online) The current on the surface of the TI induced by the rotating exchange field for $\phi=\pi/2,\ \pi/2+0.2$ and $r_0=1/\sqrt{2}$. 
The straight line denotes the full-wave rectification given by eq.~(\ref{fullwave}).
The dashed line represents the current when the rotation axis of the exchange field deviates from x-y plane. The abrupt change occurs at $\hat{M_z}=0$ and will be smoothed by the high frequency mode in the short time period $\delta t\sim \hbar/\Delta$.
\label{Fig;current1}} 
\end{center}
\end{figure}

\begin{figure}[htbp]
\begin{center}
\includegraphics[scale=0.5]{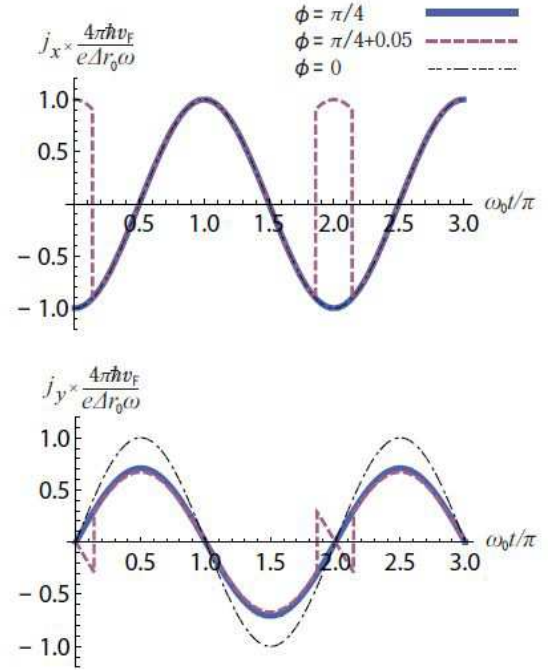}
\caption{(color online) The current for $\phi=\pi/4,\ \pi/4+0.05,0$ and $r_0=1/\sqrt{2}$. For $\phi=\pi/4$, the z-component of the rotating exchange field $\hat{M_z}$ touches zero. For $\phi>\pi/4$ the abrupt change of the current appears when $\hat{M_z}=0$.
\label{Fig;current2x}} 
\end{center}
\end{figure}

Let us examine the case of $\phi=\pi/2$ in detail, 
where the magnetization reads 
\begin{equation}
\hat{\bolM}=(r_0\sin \omega_0 t, \sqrt{1-r_0^2},-r_0\cos\omega t) \ .
\end{equation}
The mass term is thus $m=-r_0\Delta\cos \omega_0 t/v_{\text{F}}^2$ and 
eq.~(\ref{current}) reduces to $\VEV{j^0}=0$ and
\begin{equation}
\VEV{j^x}=\frac{e\Delta}{4\pi\hbar v_{\text{F}}}r_0\omega_0|\cos \omega_0 t|,\ \VEV{j^y}=0\ .
\label{fullwave}
\end{equation}
In this case, the generated current is a direct current as shown by the straight line in Fig.~\ref{Fig;current1} 
in spite of the oscillation of the sign of the magnetization. 
By using the reasonable material parameters\cite{Nomura-Nagaosa} as $v_{\text{F}}=4\times 10^{5}$ m/s, $\Delta=0.2$ eV, 
\footnote{In these parameters, the induced electric field from the precessing magnetization may be given by $r\omega_0 \times 10^{-1\sim -2}$ V/m as discussed in Ref.37, where we assume $M_z$ as $10\sim100$ mT. Then, the induced current is given by $j= r\omega_0\times10^{-6 \sim -7}$ A/m. Hence, the induced electric field can be neglected in a micrometer-size system if $\omega_0=1$GHz. This restriction on the system size will be relaxed for a smaller frequency $\omega_0$. At any rate, the induced current does not matter in observing the DC output from the rectified current since the total induced current in a period is zero. }
the current density is given by $\VEV{j^x}=9.6\times10^{-12}r_0\omega_0|\cos \omega_0 t|$ A/m.
If the frequency $\omega_0$ is 1GHz, $\VEV{j_x}=10^{-2}\sim10^{-3}$ A/m, which is large enough for detection. 
When $\phi$ deviates from $\pi/2$, i.e., $\phi=\pi/2+\delta\phi$, 
$\VEV{j^x}$ exhibit a jump between 
$\pm \frac{e\Delta \omega_0}{4\pi\hbar v_{\text{F}}} \sqrt{1-r_0^2} \tan\delta\phi$ 
($=\mp \frac{e\Delta \omega_0}{4\pi\hbar v_{\text{F}}} \sqrt{1-r_0^2} \cot\phi$)
when $m=0$, namely, when $r_0\cos\omega_0 t=-\sqrt{1-r_0^2}\tan\delta\phi$. 
A jump becomes significant when $r_0\sim 1/\sqrt{2}$ and 
$\phi$ is close to $\pi/4$ as shown by the dashed line in Fig.~\ref{Fig;current2x}.


\section{validity of adiabatic approximation}
\label{adiabatic}
We have shown above that a topological effect of the parity anomaly is directly 
observable either as a rectification effect or as a jump in the electric current. 
As we have pointed out, however, 
eq.~(\ref{current}) is not correct in the vicinity of $m=0$ due to the non-adiabaticity. 
In this section, we investigate the effect of non-adiabaticity. 
The result of eq.~(\ref{current}) is correct as far as 
$\omega_0 \ll mv_{\text{F}}^2/\hbar$ (adiabatic limit) since the gradient expansion is implied. 
In the opposite case of $mv_{\text{F}}^2/\hbar\lesssim \omega_0$, the time-dependence of the "mass" term becomes essential and a deviation from the adiabatic limit (eq.~(\ref{current})) may arise. 
We will demonstrate here explicitly that the non-adiabatic correction from the dynamic mass term 
($m(t)$) is negligibly small and that the time window for the non-adiabatic correction to be dominant is very narrow in the low frequency regime $\omega_0 \ll\Delta/\hbar$, which practically justifies the adiabatic treatment. 
For $\Delta=0.2$ eV, this condition is validated for $\omega_0 \ll400$ THz. 

\begin{figure}[htbp]
\begin{center}
\includegraphics[scale=0.5]{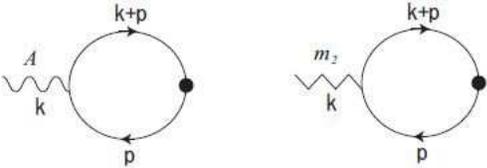}
\caption{The diagrams that contribute the current in the first order in $A$ and $m_2$.
\label{Fig;diagram}} 
\end{center}
\end{figure}

In the following calculation, we set $v_{\text{F}}=1$ and $\hbar=1$ for simplicity.
To take into account the effect of the dynamical mass, we introduce the small fluctuation, $m_2(x)$, as 
\begin{equation}
m(x)=m+m_2(x)\ ,
\end{equation}
where $m=\frac{\Delta}{TV}\int d^3x \hat{M}_z(x)$ is the average mass ($TV$ is the volume of the spacetime) and $x=(t,\bolx)$ denotes space-time coordinates. 
The mass fluctuation leads to an additional potential term in the Lagrangian:
\begin{equation}
\delta L=-m_2(x)\bar{\psi}(x)\psi(x)\ .
\end{equation}
Including the correction to the linear order in  $m_2$ (Fig.~\ref{Fig;diagram}), the current becomes
\begin{equation}
\begin{split}
\VEV{j^\mu(k)}=iA_\nu(-k)\int\frac{d^3p}{(2\pi)^3} {\rm tr}[S(k+p)\gamma^\nu S(p)\gamma^\mu]\\
+im_2(-k)\int\frac{d^3p}{(2\pi)^3} {\rm tr}[S(k+p)S(p)\gamma^\mu]\ ,\\
\end{split}
\end{equation}
where
\begin{equation}
S(k)=\frac{i(\gamma^\nu k_\nu +m)}{k^2-m^2}\ ,
\end{equation}
and $k^2=(k^0)^2-(k^1)^2-(k^2)^2$.
It is calculated as
\begin{equation}
\begin{split}
\VEV{j^\mu(k)}=&\left((g^{\mu\nu}-\frac{k^\mu k^\nu}{k^2})\Pi_1(k)+
\epsilon^{\mu\rho\nu}\Pi_{2\rho}(k)\right)A_\nu(-k)\\
&+\Pi_m^\mu(k)m_2(-k),
\end{split} \label{13}
\end{equation}
where 
\footnote{The calculation of eq.~$(\ref{FirstA})$ has been performed in many papers such as Ref.~\onlinecite{P-anomaly1} by using the various regulators, but the equation is not explicitly shown to the best of our knowledge. }
\begin{subequations}
\begin{align}
\Pi_1(k)&=\frac{k^2}{2\pi}\int_0^1dx\frac{x(1-x)}{(m^2-x(1-x)k^2)^{1/2}}\label{screening}\nonumber\\
&=O(\frac{k^2}{m})\ ,\\
\Pi_{2\rho}(k)&=-i\frac{mk_\rho}{4\pi}\int_0^1dx\frac{1}{(m^2-x(1-x)k^2)^{1/2}}\nonumber\\
&=-i\frac{mk_\rho}{4\pi|m|}+O(k_\rho\frac{k^2}{m^2})\ ,\label{chern}
\end{align}
\label{FirstA}
\end{subequations}
and
\begin{equation}
\begin{split}
\Pi_m^\mu(k)&=\frac{mk^\mu}{\pi}\int_0^1dx\frac{(1-2x)}{(m^2-x(1-x)k^2)^{1/2}}\\
( &=0\ \text{for}\ k^2 \leq 4m^2 )\ .
\label{m_2C}
\end{split}
\end{equation}
The current within the linear response theory is obtained by substituting 
$k^0\rightarrow k^0+i0^+$.
In this calculation, the ultraviolet divergence is regularized 
\footnote{In these diagrams, counterterms are not needed to remove the divergence and the renormalization is not applied.} 
by the dimensional regularization of 't Hooft and Veltman.\cite{D-regularization}
The dimensional regularization is useful in the sense that the obtained result automatically satisfies the gauge invariance of the theory.
To regularize the divergent diagram, there are various types of regulators other than the dimensional regularization. 
For example, the hardcutoff that assumes the momentum cutoff in the divergent integral is a simple regulator. 
However, as known in the gauge theory of the Dirac fermion, the resulting current from the hardcutoff does not satisfy the Ward-Takahashi identity.\cite{PeskinSchroeder} 

As seen from eqs.~(\ref{FirstA}), (\ref{m_2C}), TM term is dominant when $k^2\ll m^2$.
In the time-dependent system we consider, $k^2\sim (k^0)^2=\omega_0^2$ and thus the adiabatic condition where eq. (\ref{current}) is correct is given by $\omega_0\ll m$.
When $m\sim O(\omega_0)$, all the contributions  in eq. (\ref{13}) becomes the same order, 
$\Pi_1\sim\Pi_{2\rho}\sim \Pi \sim O(\omega_0)$ as we see from dimensional analysis.
However, the time-window for $m\lesssim \omega_0$ is $\hbar/\Delta$ and is very narrow if $\omega_0\ll \Delta/\hbar$.
The non-adiabatic contribution is therefore essentially not observable if 
$\omega_0\ll \Delta/\hbar$, 
and then the rectification effect and sudden jump in the pumped current due to the anomaly would be observable even in the non-adiabatic regime.

\section{effect of impurities}
Next, let us consider the effect of impurities on the massive Dirac fermion.
The TM is topologically protected and robust against impurities in the gapped phase in contrast to the metallic phase where impurities induce a nontrivial dynamics.\cite{TIimp} 
The effect of the impurities in the Hamiltonian is given by
\begin{equation}
H_{\rm imp}=\int d^2x v_i(\bolx)\psi^\dagger(\bolx)\psi(\bolx)\ ,
\label{impurity}
\end{equation}
where $v_i$ is the impurity potential. In this paper, we consider
$
v_i(\bolx)=\sum_n^{N_i} u\delta(\bolx-\bolx_n)
$, 
where $u$ is the strength of the impurity scatterers, $N_i$ is the number of the impurities and $\bolx_n$ is the position of the $n$-th impurity. We will average over the positions of the impurities as
$\VEV{v_i(\bolx)}_i =0$
and
$\VEV{v_i(\bolx_1)v_i(\bolx_2)}_i =  \frac{N_i u^2}{V^2} \delta(\bolx_1-\bolx_2)$,
where $V$ is the volume of the space. 
For treating the dissipation driven by the impurities, we use the Keldysh Green's function. 
The current is given by
\begin{equation}
\VEV{j^i(x)}=ie\hbar v_{\text{F}}
\epsilon_{ij}\text{tr}[\sigma^j G^{<}(x,x)]\ ,
\end{equation}
where $i,j=(x,y)$ 
and $G^{<}$ is the lesser
Green's function,
$G^<_{\sigma,\sigma^\prime}(x,x^\prime) =\frac{i}{\hbar}\VEV{\psi^\dagger_{\sigma^\prime}(x^\prime)\psi_\sigma(x)}$.
In the first order in $A$, the current reads
\begin{align}
\VEV{j^i(x)}
=
&\frac{ie^2\hbar v_{\text{F}}}{V}
\epsilon_{ij} \epsilon_{ab}
\int\frac{d^3k}{(2\pi)^3} \int\frac{d^3p}{(2\pi)^3} e^{ipx} A^a(p)
\nonumber \\
&\times
[ f(k_0+\frac{p_0}{2}) - f(k_0-\frac{p_0}{2}) ]
\nonumber \\
&\times \text{tr}
[ \sigma^j g^r(k-\frac{p}{2}) \sigma^b g^a(k+\frac{p}{2}) ]\ ,
\end{align}
where $f$ is the Fermi distribution function. 
Here the advanced (retarded) Green's function, $g^a$ [$=(g^r)^*$], is given as
\begin{equation}
g^a(k)
=
\frac{(\hbar k_0-i\eta){\bf 1}_{2\times 2}+mv_{\text{F}}^2\sigma^z-\hbar v_{\text{F}}(k_x\sigma_y-k_y\sigma_x)}{(\hbar k_0-i\eta)^2-(mv_{\text{F}}^2)^2-(\hbar v_{\text{F}}|\bolk|)^2}\ ,
\end{equation}
where $\eta$ represents the self energy of the fermion.\cite{Keldysh} 
To see the effect of the impurities on TM, we focus on the term proportional to
$\text{tr}[\sigma^\mu\sigma^\nu\sigma^\rho]=2i\epsilon^{\mu\nu\rho}$.
As a result, it is found that TM term does not change even if the impurities exist. 
Now, let us discuss the effect of the higher order diagrams in the context of the Coleman-Hill theorem.\cite{Coleman-Hill} 
This theorem states that the correction from scalar, spinor, vector and gauge fields does not affect TM in the photon self energy appearing in the 1-loop diagram, if the Ward-Takahashi identity is satisfied, the momentum is conserved and the Green's function is massive. 
By averaging over the positions of impurities, the impurity can be viewed as the chargeless scalar field, and then the Ward-Takahashi identity and the momentum conservation is satisfied. 
Because of the averaging process, the impurity by itself does not induce the current ($\VEV{v_i(\bolx)}_i =0$), in contrast to the $m_2$ term discussed in the previous section.
Therefore, the Coleman-Hill theorem\cite{Coleman-Hill} can be applied to our calculation and the impurities do not affect TM at any order of loops.

\section{conclusion and discussion}
We have discussed the current dynamics on the surface of the TI by the precessing magnetization. The rectification effect and the jump of the current have been found to arise from the parity anomaly in the adiabatic regime. We have discussed that the effect of the non-adiabaticity can be neglected. 
To remove the divergence in the diagram preserving the gauge invariance of the theory, the dimensional regularization\cite{D-regularization} is used. It should be pointed out that the dimensional regularization can be also applied to the calculation of other physical quantities such as the conductivity in the metallic phase of the TI and hence would be a useful tool to investigate properties of TI. Also, the robustness of the TM against the impurities is reconfirmed. 

Finally, let us discuss the possibility of a rectification driven by a light. In our study, it is essential for the rectification that the mass term controls the sign of the Hall current. 
The same goes for the case of a circularly polarized light radiation to the surface of a bare TI, and thus a rectification is expected.
In this case, the light produces a varying mass term $m=g\mu_B B_z(\bolx)/v_{\text F}^2$ by the Zeeman coupling, where $B_z$ is the magnetic field of light. The Zeeman couplings to $B_x,B_y$ does not produce an observable current since these are much weaker than the exchange coupling discussed in Sec.~\ref{Sec:rect}. 
On the other hand, the rotating electric field of light induces the current $j_{i}=\frac{e^2}{2h}\frac{m}{|m|}\epsilon_{ij}E_{j}$ with $i,j=(x,y)$. If the pointing vector is within the surface plane $({\bf B}\times{\bf E})_z=0$, $E_{x}$ and $E_y$ change the sign when $B_z=0$; the rectification occurs. 
For experimental realization, a strong laser would be necessary to satisfy the adiabatic condition $\hbar \omega, \hbar v_{\text F} k\ll g\mu_B |B_z|$. 
In addition, the fine tuning of the Fermi level at the Dirac point needs to be carried out precisely, since the induced gap is assumed to be small, for example, $10^{-5}$ eV for $B_z=100$ mT. 

We expect that the nontrivial-current dynamics driven by the quantum anomaly can be applied to new spintronics devices.

{\it Note added}. After the completion of our work, we became
aware of the paper by B.~D\'{o}ra {\it et al}., who studied
a rectification effect on a edge state of 2D TI induced by a circularly polarized light both in the adiabatic and non-adiabatic regime.\cite{1Drect} 
The edge state is described by the (1+1)D-Dirac fermion. 
In their setting, the rectification in the adiabatic regime is driven by the topological excitation\cite{review,GoldstoneWilczek} related to the Dirac mass term.
On the other hand, in our (2+1)D study, the parity anomaly plays an essential role in the rectification. 

\begin{acknowledgements}
We thank S. Murakami and E. Saitoh for discussions. 
G.T. was supported by a Grant-in-Aid for Scientific Research (B) 
(Grant No. 22340104) from the Japan Society for the Promotion of Science 
and the UK-Japanese Collaboration on Current-Driven Domain Wall Dynamics from JST. 
T.Y. was supported by Grant-in-Aid for Young Scientists (B) (No. 23740236) and the ``Topological Quantum Phenomena" (No. 23103505) Grant-in Aid for Scientific Research on Innovative Areas from the Ministry of Education, Culture, Sports, Science and Technology (MEXT) of Japan. 

\end{acknowledgements}



\begin{thebibliography}{101}
\bibitem{Moore} J. E. Moore and L. Balents, Phys. Rev. B \textbf{75}, 121306(R) (2007).

\bibitem{Fu} L. Fu, C. L. Kane, and E. J. Mele, Phys. Rev. Lett. \textbf{98}, 106803 (2007).

\bibitem{ME}
X.-L.~Qi, T.~L.~Hughes, and S.-C.~Zhang, 
Phys. Rev. B {\bf 78}, 195424 (2008). 

\bibitem{Hasan} M. Z. Hasan and C. L. Kane, Rev. Mod. Phys. \textbf{82}, 3045 (2010).

\bibitem{review}
X.-L.~Qi and S.-C.~Zhang, 
Rev. Mod. Phys. {\bf 83}, 1057 (2011).
\bibitem{monopole}
X.-L.~Qi, R.~Li, J.~Zang, S.-C.~Zhang, Science {\bf 323}, 1184
(2009).

\bibitem{P-anomaly1}
A.~N.~Redlich, 
Phys. Rev. D {\bf 29}, 2366 (1984).
\bibitem{P-anomaly2}
A.~J.~Niemi and G.~W.~Semenoff, 
Phys. Rev. Lett. {\bf 51}, 2077 (1983). 
\bibitem{Tmass}
S.~Deser, R.~Jackiw and S.~Templeton, 
Phys.~Rev.~Lett. {\bf 48}, 975 (1982). 
\bibitem{TKNN}
D.~J.~Thouless, M.~Kohmoto, M.~P.~Nightingale and M.~den~Nijs,
Phys. Rev. Lett. {\bf 49}, 405 (1982).
\bibitem{Coleman-Hill}
S.~Coleman and B.~Hill, Phys. Lett. B {\bf 159}, 184 (1985).
\bibitem{anomaly}
S.~Adler, Phys. Rev. {\bf 177}, 2426 (1969);
J.~Bell and R.~Jackiw, Nuovo Cimento {\bf 60A}, 47 (1969).
\bibitem{PeskinSchroeder}
M.~E.~Peskin and D.V.~Schroeder, 
{\it An Introduction to Quantum Field Theory}, 
(Westview press, 1995)
\bibitem{Haldane}
F.~D.~M. Haldane, 
Phys. Rev. Lett. {\bf 61}, 2015 (1988). 
\bibitem{KaneMele}
C.~L.~Kane and E.~J.~Mele, 
Phys. Rev. Lett. {\bf 95}, 226801 (2005). 
\bibitem{Zhang} S.~C.~Zhang, Int. J. Mod. Phys. B \textbf{6}, 25 (1992).

\bibitem{XGWenText}
X.~G.~Wen, 
{\it Quantum Field Theory of Many-Body Systems}, 
(Oxford university press, 2004).
\bibitem{Nomura-Nagaosa2}
K.~Nomura and N.~Nagaosa, 
Phys. Rev. Lett. {\bf 106}, 166802 (2011). 
\bibitem{Inverse}
I.~Garate and M.~Franz, 
Phys.~Rev.~Lett. {\bf 104}, 146802 (2010). 

\bibitem{Tse} W.-K.~Tse and A.~H.~MacDonald, Phys. Rev. Lett. \textbf{105}, 057401 (2010).

\bibitem{Maciejko} J.~Maciejko, X.-L.~Qi, H.~D.~Drew, and S.-C. Zhang, Phys. Rev. Lett. \textbf{105}, 166803 (2010).

\bibitem{Yokoyama1} T.~Yokoyama, Y.~Tanaka, and N.~Nagaosa, Phys. Rev. B \textbf{81}, 121401(R) (2010).

\bibitem{Yokoyama2} T.~Yokoyama, J.~Zang, and N.~Nagaosa, Phys. Rev. B \textbf{81}, 241410(R) (2010).

\bibitem{Burkov} A.~A.~Burkov and D.~G.~Hawthorn, Phys. Rev. Lett. \textbf{105}, 066802 (2010).

\bibitem{Nomura-Nagaosa}
K.~Nomura and N.~Nagaosa, 
Phys. Rev. B {\bf 82}, 161401 (2010).

\bibitem{Yokoyama3} T.~Yokoyama and S.~Murakami Phys. Rev. B \textbf{83}, 161407(R) (2011) 

\bibitem{Yokoyama4} T.~Yokoyama, Phys. Rev. B \textbf{84}, 113407 (2011).

\bibitem{Mahfouzi}
F.~Mahfouzi, N.~Nagaosa and B.~K.~Nikoli\'{c}, 
arXiv:1112.2314.

\bibitem{Tserkovnyak}
Y.~Tserkovnyak and D.~Loss, 
arXiv:1112.5884.

\bibitem{D-regularization}
G.~'t~Hooft and M.~J.~T.~Veltman, Nucl. Phys. B {\bf 44}, 189 (1972);
For review, see Sec.~7.5 in Ref.~\onlinecite{PeskinSchroeder}.
\bibitem{grapheneDR}
V.~Juri$\check{\text{c}}$i$\acute{\text{c}}$, O.~Vafek, and I.~F.~Herbut, 
Phys. Rev. B {\bf 82}, 235402 (2010).
\bibitem{TIimp}
P.~Schwab, R.~Raimondi and C.~Gorini, Eur. Phys. Lett. {\bf 93}, 67004 (2011);
D.~Culcer and S.~D.~Sarma, Phys. Rev. B {\bf 83}, 245441 (2011). 
\bibitem{Keldysh}
For the detailed formulation of the Keldysh Green's function, 
see H.~Haug and A.-P.~Jauho, 
{\it Quantum Kinetics in Transport and Optics of Semiconductors} 
(Springers, 2007). 
\bibitem{1Drect}
B.~D\'{o}ra, J.~Cayssol, F.~Simon and R.~Moessner, 
Phys. Rev. Lett. {\bf 108}, 056602 (2012).
\bibitem{GoldstoneWilczek}
J.~Goldstone and F.~Wilczek, 
Phys.~Rev.~Lett, {\bf 47}, 986 (1981).  
\end{thebibliography}
\end{document}